\documentclass[12pt]{iopart}
\usepackage{graphicx}
\begin{document}
\title[Radiative Pulsed L-Mode ARC]{Radiative pulsed L-mode operation in ARC-class reactors}

\author{S J Frank$^{1\dag}$, C J Perks$^1$, A O Nelson$^2$, T Qian$^3$, S Jin$^3$, A Cavallaro$^1$, A Rutkowski$^3$, A Reiman$^3$, J P Freidberg$^1$, P Rodriguez-Fernandez$^1$, D Whyte$^1$}
\date{May 2022}

\address{$^1$ Plasma Science and Fusion Center, Massachusetts Institute of Technology, 
Cambridge, MA 02139, USA \\
$^2$ Department of Applied Physics and Applied Mathematics, Columbia University, New York, NY 10027, USA \\
$^3$ Princeton Plasma Physics Laboratory, Princeton, NJ 08540, USA\\
$\dag$ Corresponding author: frank@psfc.mit.edu\\
}

\begin{abstract}
    A new ARC-class, highly-radiative, pulsed, L-mode, burning plasma scenario is developed and evaluated as a candidate for future tokamak reactors. Pulsed inductive operation alleviates the stringent current drive requirements of steady-state reactors, and operation in L-mode affords ELM-free access to $\sim90\%$ core radiation fractions, significantly reducing the divertor power handling requirements. In this configuration the fusion power density can be maximized despite L-mode confinement by utilizing high-field to increase plasma densities and current. This allows us to obtain high gain in robust scenarios in compact devices with $P_\mathrm{fus} > 1000\,$MW despite low confinement. We demonstrate the feasibility of such scenarios here; first by showing that they avoid violating 0-D tokamak limits, and then by performing self-consistent integrated simulations of flattop operation including neoclassical and turbulent transport, magnetic equilibrium, and RF current drive models. Finally we examine the potential effect of introducing negative triangularity with a 0-D model. Our results show high-field radiative pulsed L-mode scenarios are a promising alternative to the typical steady state advanced tokamak scenarios which have dominated tokamak reactor development.
\end{abstract}

\submitto{\NF}
\ioptwocol

\maketitle

\section{Introduction}
For fusion to occupy a share of the global energy production mix by the mid-21st century, a physically viable and economically attractive fusion pilot plant should be designed by the late 2020s. Development of a pilot plant plan on this short timeline is one of the key goals outlined in recent reports regarding the future of fusion energy from both the National Academy of Sciences, Engineering, and Medicine and the Fusion Energy Sciences Advisory Committee \cite{FESAC,NASEM}. Several potential pathways to a pilot plant have  been established. Most of these strategies aim to achieve reactor-relevant core conditions using the advanced tokamak (AT) regime characterized by H-Mode operation (or I-mode in the case of ARC) \cite{Wagner1984, Sorbom2015} augmented by internal transport barriers to further enhance confinement. Enhanced confinement and internal transport barriers create large pressure gradients providing significant bootstrap current fractions, $f_{bs}$. High confinement time then, allows minimization of plasma current improving stability while also reducing the external current drive requirements needed to obtain a steady-state reactor \cite{Najmabadi2006, Sorbom2015, Buttery2021}. 

Recently, high temperature superconductor (HTS) technology has dramatically increased the achievable on-axis magnetic field in reactor designs \cite{Mangiarotti2015, Michael2017}. Since fusion power density scales like $P_\mathrm{fus}\sim \beta^2 B^4$ \cite{Wesson}, this technological advancement provides opportunities to develop or improve self-consistent reactor scenarios \cite{Zohm2010}. AT scenarios, combined with the stronger on-axis magnetic fields available using HTS have been shown to significantly decrease the minimum device major radius needed in a steady state reactor \cite{Olynyk2012,Sorbom2015,Buttery2021}. As size is one of the largest drivers increasing levelized cost of electricity, high $B$ scenarios are economically attractive \cite{Sorbom2015,Zohm2019, Wade2021}.

\begin{figure}
	\includegraphics[width=1\linewidth]{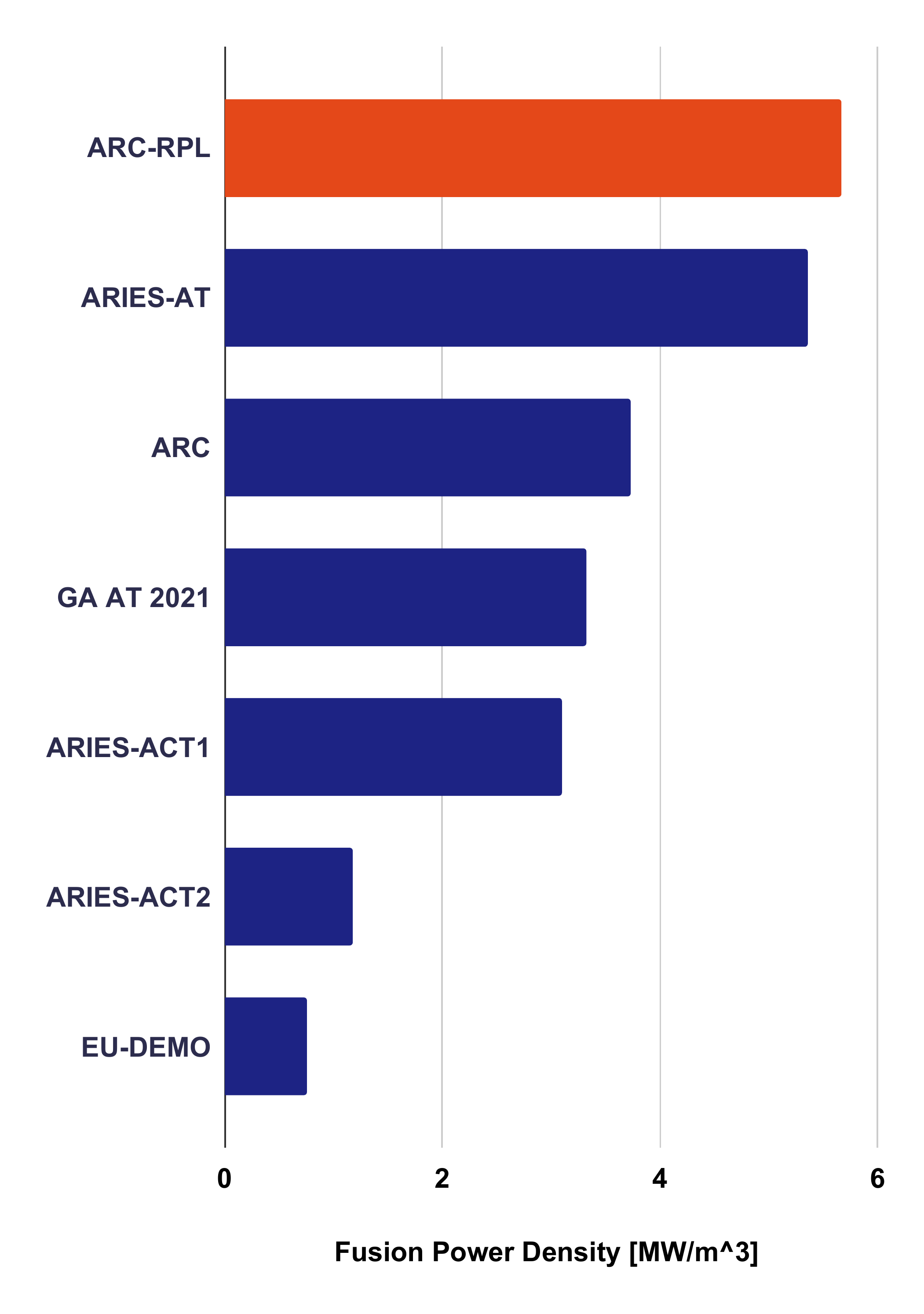}
	\caption{A fusion power density (fusion power $P_{fus}$ over the reactor volume $V$, or $P_{fus}/V$) comparison between the RPL mode ARC concept analyzed in this work and a number of other proposed tokamak reactor designs \cite{Najmabadi2006,Kessel2015,Federici2014,Sorbom2015,Buttery2021}.} 
	\label{fig:powdencomp}
\end{figure}

While the benefits of high $B$ in the AT regime are now widely recognized, it is worth considering whether strong on-axis magnetic fields can enable access to pilot plant designs \textit{outside} of typical AT scenarios. In this work, we explore one such class of reactor scenarios enabled by HTS magnets. We propose a pulsed L-mode operation scenario with largely radiative heat exhaust by impurity line radiation in the plasma edge, denoted from now on as ``RPL-mode," in an ARC-class high-field compact tokamak ($R\sim $ 3 m to 4 m, $a\sim1$ m, $B \sim 10$ T). The RPL-mode facilitates power density maximization while removing the need for an advanced divertor by dramatically decreasing the power exhausted to the scrape-off layer (SOL). This allows us to produce some of the highest power densities of any tokamak reactor concept, demonstrated in Fig.~\ref{fig:powdencomp}. 

Here we demonstrate the functional viability of high-field RPL-modes during flattop and establish a physics basis for their consideration in reactor design. To do so, first we review the basic physics of radiative L-mode reactors and historical experiments on radiative L-modes in Sec.~\ref{sec:background}. We develop a 0-D RPL-mode design-point, obtained using well-established physics models, to extrapolate L-mode physics to reactor-relevant conditions in Sec.~\ref{sec:0D}. Then, starting from the 0-D design point we perform coupled self-consistent simulations of the MHD equilibrium, turbulent transport, and RF heating to refine our RPL-mode scenario and determine its operational viability in Sec.~\ref{sec:integrated}. Next, in Sec.~\ref{sec:negD} we examine negative triangularity RPL-modes using a modified version of the 0-D model from Sec.~\ref{sec:0D}. Finally, we discuss reactor engineering considerations associated with the development of RPL-mode reactors.

\section{The Physical Basis For RPL-Modes}
\label{sec:background}

In this section, we argue that L-mode operation provides a possible heat exhaust solution for compact high field reactors. The compact high-field pathway to tokamak fusion reactors results from balancing tokamak operational limits to maximize benefits from large on-axis magnetic fields. Tokamak confinement, as prescribed by energy confinement scaling laws, is only weakly affected by increasing toroidal magnetic field. For example, the ITER-89P and ITER-98Py2 energy confinement scaling laws have only weak dependencies on the imposed toroidal magnetic field, $\sim B^{0.2}$ and $\sim B^{0.15}$, respectively \cite{ITER1999,Doyle2007}. However, increasing $B$ increases allowable plasma current while keeping important magneto-hydrodynamic (MHD) stability parameters, notably the edge safety factor $q^*$, constant \cite{Freidberg2014}. For an elliptical equilibrium,
\begin{equation}
q^* \simeq \frac{\pi a^2 B_0 (\kappa^2+1)}{ \mu_0 R_0 I_p},
\end{equation}
where $a$ is the plasma minor radius, $B_0$ is the on-axis toroidal field, $\kappa$ the elongation, and $R_0$ the major radius. To ensure stability against external kinks, $q^* > 2$ is required. The kink limit sets an easily calculable 0-D absolute upper limit on $I_p$, which increases linearly as $B$ is increased. Increasing $I_p$ increases confinement (energy confinement time, $\tau_E \propto I_p$) and the maximum achievable line averaged density $n_g$ prescribed by the Greenwald limit \cite{Greenwald2002}:
\begin{equation}
n_g [10^{20} m^{-3}] = \frac{I_p [\textrm{MA}]}{\pi (a [\textrm{m}])^2}.
\end{equation}
Fusion power density in a DT-fueled fusion device with 50/50 fuel mix goes:
\begin{equation}
    \mathcal{P}_\mathrm{fus} = 1/4\,n^2 \langle\sigma v\rangle_{DT} E_{DT},
\end{equation}
where $\langle\sigma v\rangle_{DT}$ is the Maxwell averaged DT fusion reaction rate and $E_{DT}$ is the energy produced by each DT fusion reaction (17.6 MeV total, a 3.5 MeV alpha and a 14.1 MeV neutron). High on-axis field increases both the allowable current $I_p$ and density $n$ and enabling high fusion power density even with low confinement. As the Greenwald density is also increased by small $a$, power density is further maximized by using more compact devices.

There are now a number of tokamak designs that leverage HTS to achieve high fusion power densities in compact devices \cite{Olynyk2012, Sorbom2015, Creely2020}. However, energy density in high-field compact devices is limited by the available heat exhaust solutions, and advanced divertors are required to minimize damage to plasma-facing components (PFC) \cite{Kuang2018}. Introducing radiative heat exhaust allows us to push past divertor-related energy density limits and maximize fusion power density. Unlike heat exhaust through the divertor, where only a small physical area is exposed to heat loading, radiative heat exhaust uniformly distributes the heat load over the plasma viewing PFC increasing the exposed area. Large quantities of radiative heat exhaust can be used to substantially reduce the peak energy fluences by reducing heat flux to the divertor. The amount of radiative exhaust is characterized by, $f_{rad}=P_{rad}/P_{exhaust}$ the radiated power fraction. In this fraction, $P_{exhaust}$ is defined by power balance:
\begin{equation}
    \label{eq:powbal}
    \eqalign{P_{exhaust} = P_{SOL} + P_{rad} \cr = W_{tot}/\tau_{E} = P_{fus,\alpha}+P_{aux}},
\end{equation}
where $W_{tot}$ is the total stored energy, $P_{SOL}$ is the power convected through the scrape-off layer, $P_{fus,\alpha}$ is the power from fusion alphas, $P_{rad}$ is the total power radiated within the last closed flux surface, and $P_{aux}$ is the total auxiliary power used for heating and current drive. Higher $f_{rad}$, is almost always desirable as it reduces the peak heat loads on the divertor. EU-DEMO, for example, is expected to require core radiation fractions $> 60 \%$ \cite{Zohm2013} and  proposed compact reactors are also expected to have radiation fractions of $\sim 40-60 \%$ \cite{Kuang2018,Buttery2021}. While current experiments obtain high $f_{rad}$ with intrinsic low-Z or medium-Z impurities \cite{Greenwald1997}, intentional puffing or pellet injection of highly efficient radiators, such as Kr and Xe, using feedback control will be necessary to obtain large $f_{rad}$ without excessive dilution \cite{Kallenbach2013}. 

Maintaining high $f_{rad}$ in H-mode reactor scenarios requires careful management of radiated power to avoid an H-L back-transition. If the power $P_\mathrm{SOL}$, transported across the separatrix with surface area $S$, falls below the H-L transition power \cite{Martin2008}:
\begin{equation}
    \label{eq:P_LH}
    \eqalign{P_\mathrm{LH} \, [\textrm{MW}] = 0.049\, (\bar{n}_e [10^{20} \, \textrm{m}^{-3}])^{0.72} \cr \times (B [\textrm{T}])^{0.8} (S [\textrm{m}^2])^{0.94}},
\end{equation}
the reactor will leave H-mode. The resulting large drop in confinement typically causes the plasma to radiatively collapse and disrupt. The minimum divertor heat loading requirement in enhanced confinement is set by the need to maintain a sufficient $P_{SOL} \ge P_{LH}$. The transition power requirement is problematic as maintaining $P_{SOL} \ge P_{LH}$ necessitates the use of advanced divertors in compact pilot plants increasing their complexity \cite{Kuang2018}. Simply increasing device size does not necessarily solve the heat exhaust problem. If $R$ is increased, the ITER98y2 scaling \cite{ITER1999,Doyle2007} suggests fusion power will increase $\propto R^{1.97}$, and $P_{LH}$ will increase for fixed aspect ratio $\propto R$ (assuming D-T operation and ion temperatures $T_i \sim 15-25$ keV such that the stored energy roughly scales with the fusion power \cite{Wesson}). Analysis using the Eich scaling \cite{Eich2013} suggests divertor power handling scales $\propto R$ for fixed $B$ \cite{Labombard2015}, but moving to larger R at fixed aspect ratio reduces $n_g$ and the allowable operational density making dissipative divertor solutions more difficult to implement. This simplified analysis suggests that in ELMy H-mode increasing $R$ increases fusion power faster than the divertor's ability to exhaust it. Therefore, fusion power density will need to be limited in large devices due to divertor constraints. This means increasing device size does not necessarily solve our heat exhaust problems unless the fusion power output is intentionally reduced.

While the applicability of radiative heat exhaust in H-mode is limited by constraints imposed by the L-H transition power $P_{LH}$, high field magnets allow us to consider radiative \textit{L-mode} reactor scenarios. Previously, L-mode scenarios were not reactor-relevant due to lack of sufficient confinement, but high $B$ from HTS magnets enables $I_p$ and $n$ maximization, substantially increasing the achievable fusion power with L-mode confinement. Such $n$ maximization could be difficult in H-modes as, at densities corresponding to 0.8-1.0 $n_g$, H-mode access can become impaired and confinement degradation occurs (this is sometimes referred to in the literature as an H-mode density limit) \cite{Huber2013,Bernert2014}. Radiative L-modes have long been recognized as a possible solution to the heat exhaust problem \cite{Gibson1977,Lackner1984}, and since L-modes do not have a power threshold they can operate with very large $f_{rad}$ and $f_g$.

Radiative heat exhaust was examined experimentally in a number of devices. Highly radiative L-mode tokamak plasmas have been demonstrated in: ISX-B, ASDEX, TEXTOR94, DIII-D, Alcator C-Mod, TFTR, ASDEX-U, and COMPASS \cite{Samm1991,Samm1993,Messiaen1994,Messiaen1996,Messiaen1997,Ongena1999,Lazarus1984,Lazarus1985,Bessenrodt-Weberpals1991,Bessenrodt-Weberpals1992,Greenwald1997,Hill1999NF,Hill1999PoP,Komm2021,Fable2021}, and in some cases these plasmas demonstrated performance enhancement above the ITER89P L-mode scaling due to ion temperature gradient (ITG) driven transport suppression from increased effective ion charge, $Z_{eff}$ \cite{Dominguez1993,Tokar1999-1,Tokar1999-2}. Of particular note, TFTR obtained large $f_{rad}$ in high-heating power discharges using Kr and Xe. These high-temperature reactor-relevant discharges with $f_{rad}$ up to $~\sim80$\% were achieved without excessive fuel dilution, radiative collapse, or energy confinement degradation. The TFTR radiative L-mode discharges demonstrated that with the proper technology, radiative L-modes could perhaps scale to reactor-relevant regimes \cite{Hill1999NF,Hill1999PoP}.
    
 However, steady state operation of radiative L-mode ARC-class reactors with $B_0 > 10$ T and $\overline{n}_e  > 2\times10^{20}$ m$^{-3}$ is anticipated to be problematic. These plasma scenarios have substantial external heating and current drive constraints, imposed by both technological limitations on radiofrequency (RF) heating sources and physical limitations on wave and neutral beam accessibility. Core neutral beam access at presently available beam energies will be poor due to the high $n$, and gyrotron sources cannot deliver the required frequencies for electron cyclotron heating (ECH) due to high $B$. Therefore, the ECH and beam systems found in other reactor studies \cite{Federici2018,Buttery2021} will not work in ARC-class devices. When the constraints presented by ARC-class devices are taken into account the following options for RF current drive and heating are: lower hybrid current drive (LHCD), ion-cyclotron heating (ICH), and fast wave current drive (including low frequency fast wave and helicons). RF current drive, however, has an unfavorable efficiency scaling \cite{Fisch1980}:
\begin{equation}\label{eq:cdeff}
    \eta_{20,CD} = \frac{I_{CD} n_{20,e} R_0}{P_{RF}} = \frac{31}{\ln \Lambda}\frac{4}{(5+Z_{eff})}\frac{1}{n_\parallel^2},
\end{equation}
where $\eta_{20,CD}$ is the current drive efficiency normalized by density and major radius, $I_{CD}$ is the driven current, $P_{RF}$ is the applied RF power, $n_\parallel$ is the parallel refractive index of the RF wave, and $\ln \Lambda \sim 17$ is the Coulomb logarithm. This scaling says that current drive efficiency $\propto 1/n_e$, which is problematic for reactor-relevant radiative L-modes as they rely on current \textit{and} density maximization to achieve large $P_{fus}$. 

In fact, reactor-relevant radiative L-modes are effectively incompatible with non-inductive operation. Nearly all of the generated electric power would need to be recirculated to the RF current drive system. We can show this by plugging some example L-mode reactor parameters into (\ref{eq:cdeff}) to estimate the RF current drive requirement: $I_p = 15$ MA, $n_{20} = 4$, $R_0 = 4$, $n_\parallel=1.75$, $Z_{eff} = 2$, and bootstrap fraction $f_{bs} = 1 - I_{CD}/ I_p = 0.3$. This gives $\eta_{CD,20} \sim 0.34$ and required RF current drive power $P_{RF} \sim 480$ MW. For an L-mode reactor producing 1500 MWth, after applying electrical conversion efficiency $\eta_e \sim 0.5$ and RF source efficiencies $\eta_{RF} \sim 0.6$ there would be effectively zero net electrical output. Accounting for wave accessibility constraints and current profile control is likely to further increase the required $P_{RF}$. 

Radiative L-modes will need to be pulsed using inductive current drive which does not suffer from RF current drive's unfavorable $1/n_e$ efficiency scaling. This completes the final piece of our physical basis for the radiative pulsed L-mode (RPL-mode). Inductive current drive of course results in finite pulse time, $\tau_{pulse}$. A high-flux HTS solenoid could provide an attractive pathway to long pulse inductive reactor operation, and such solenoids have already been identified as potentially high yield technological developments for EU-DEMO and SPARC \cite{Wesche2018,Sarasola2020,Creely2020}. Estimation of pulse time will not be attempted here. To accurately estimate pulse time requires both an engineering design for the central solenoid and a startup model that can accurately replicate the current ramp process. We have limited our scope here to flat-top physics and will focus only on the flattop plasma loop voltage, $V_{loop} \propto 1/\tau_{pulse}$. We can estimate the required $V_{loop}$ for a reactor by the following argument. It is foreseeable that longer pulse times, $\mathcal{O}(1000+)$ seconds will be desirable as they prolong the time until fatigue induced failure from cyclic loading and increase the capacity factor of a future fusion power plant. Assuming an HTS solenoid capable of generating $\sim 300$ Vs flux swings (similar to the ITER solenoid \cite{Schultz2005}) can be constructed, and roughly half of the available flux is used during startup (a reasonable assumption based on pulsed DEMO studies \cite{Federici2018}), $V_{loop} < 0.2$ V will be required during flattop.

While we will not use RF for steady-state current drive in RPL-modes, RF may have great utility as a localized control actuator. RPL-modes, particularly in enhanced confinement L-modes like those obtained with negative triangularity, can reach normalized beta values ($\beta_N \sim 2$) and significant bootstrap fractions ($f_{bs} \ge 0.25$), making them susceptible to neoclassical tearing modes. The risk of tearing modes is compounded by the large amount of radiation in the outer portion of the plasma which can approach the $q=2$ surface. The line-radiation in the outer portion of the plasma will drive a locally negative power balance that can destabilize tearing modes, reducing tokamak performance and possibly inducing a disruption. Tearing mode suppression using electron cyclotron current drive is likely not viable in an RPL-mode ARC-class reactor due to gyrotron frequency limitations, but LHCD may be a viable alternative. LHCD suppression of tearing modes has been demonstrated experimentally \cite{Warrick2000}. LHCD can also produce localized current drive in response to the temperature perturbations found in magnetic islands resultant from tearing \cite{Frank2020} and may exploit the current condensation effect to stabilize tearing modes at only modest applied RF powers \cite{Reiman2018,Jin2021,Rieman2021}. In Sec.~\ref{sec:integrated} we will demonstrate that RPL-modes are compatible with LHCD systems that can suppress 2-1 islands resultant from tearing by demonstrating an LHCD system design that locally heats the $q=2$ flux surface.      

\begin{figure*}
	\includegraphics[width=1\linewidth]{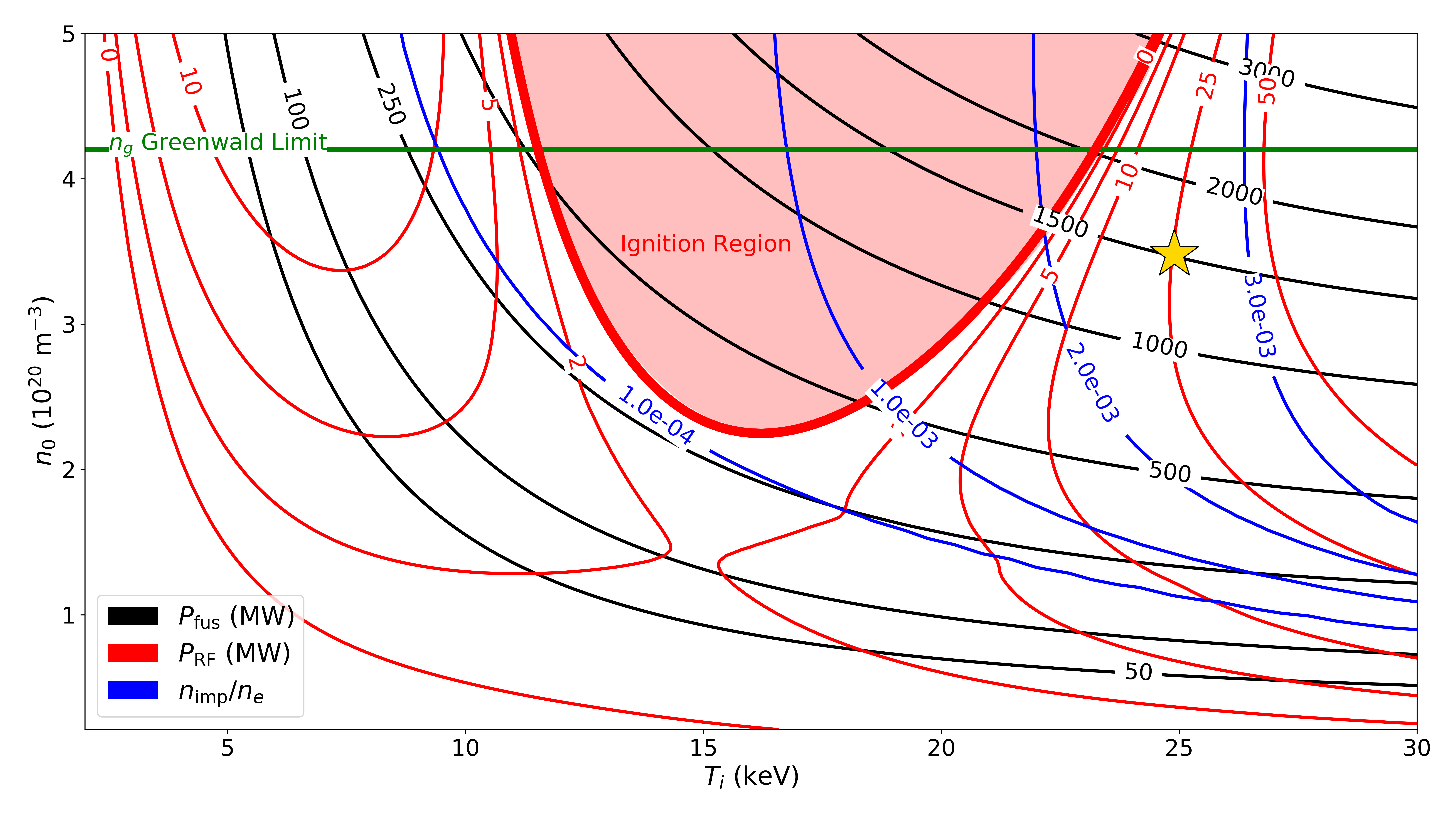}
	\caption{RPL-mode POPCONs. The axes $n_{20}$ and $T_{i}$ show the on-axis values of density in $10^{20} \, m^{-3}$ and temperature in $keV$. The operating point is denoted with the yellow star, red contour lines denote the total external heating power $P_{aux}$ (not including ohmic heating), the ignition region is shaded red, black contours denote the total D-T fusion power $P_{fus}$, and blue contours denote the impurity fraction required to maintain constant $P_{sol}=50 MW$.} 
	\label{fig:popcon}
\end{figure*}

\section{Establishing An RPL-Mode Operating Points With A 0-D Model}
\label{sec:0D}

\begin{table*}
\caption{\label{tbl:0dparams} 0-D Parameters Of ARC RPL-Mode Scenario}
\begin{tabular}{lr}
\hline
Parameter\\
\hline
Major Radius (R) & 4.2 m \\
Minor Radius (a) & 1.2 m \\
Plasma Current $\left(I_p\right)$ & 14 MA \\
Magnetic Field $\left(B_0\right)$ & 11.5 T \\
Elongation $\left(\kappa\right)$ & 1.6 \\
Confinement Enhancement Factor $\left(H_{89}\right)$ & 1.3 \\
Temp. Profile Factor $\left(\alpha_T\right)$ & 1.5 \\
Dens. Profile Factor $\left(\alpha_n\right)$ & 1.1 \\
Kink Safety Factor $\left(q^*\right)$ & 2.5 \\
Temperature on Axis ($T_{i0}$) & 24.9 keV \\
Greenwald Fraction $\left(f_g\right)$ & 0.83 \\
Fusion Power $\left(P_{fus} \right)$ & 1500 MW \\
Power Conducted to SOL $\left(P_{SOL}\right)$ & 50 MW \\
First Wall Power Loading & $\sim0.95$ MW/m$^2$\\
\hline
\end{tabular}
\end{table*}

To begin our RPL-mode scoping we performed a 0-D power balance calculation using a slightly modified version of the Plasma Operational Contours (POPCONs) technique \cite{Houlberg1982}. We assumed an elliptical core plasma with elongation $\kappa$, minor radius $a$, and major radius $R$, and imposed 1-D radial profiles of density and temperature of form $A_{sep}+(A_0-A_{sep})*(1-\rho^2)^{\alpha_A}$, where $A_{sep}$ is the parameter's value at the separatrix, $A_0$ is the parameter's value on axis, and $\rho$ is the radial coordinate which goes from 0 to 1. Densities were assumed to be $n_D = n_T = 0.5n_e f_{dil}$ where $f_{dil} = n_i/n_e$ is the ion dilution fraction. The separatrix values for these profiles are modeled after C-Mod experiments and defined to be $T_\mathrm{sep}=0.1$\,keV and $n_\mathrm{sep}=0.25n_\mathrm{0}$, where $n_0$ is the on-axis density, mimicking the SOL behavior of L-mode in Alcator C-Mod \cite{Greenwald1997,Schmidt2011}. Finally, $\tau_\mathrm{E}$ is defined using the ITER-89 L-mode power law scaling \cite{Yushmanov1990, Tsunematsu1992} multiplied by confinement enhancement factor $H_{89}$:

\begin{equation}
    \eqalign{\tau_e = 3.8\times10^4 H_{89} I_p^{0.85} B^{0.2} n^{0.1} \cr \times P_{exhaust}^{-0.5} R^{1.5} \kappa^{0.5} \epsilon^{0.3} M^{0.5}.}
\end{equation}

To obtain a self-consistent radiative boundary and ensure we maintained L-mode operation, we controlled $P_{rad}$ by varying the injected impurity fraction. $P_{rad}$ in the presence of impurities was calculated assuming ionization equilibrium with radiation rates and charge states of the impurity as a function of $T$ taken from compact polynomial fits based on the most recent Atomic Data and Analysis Structure database \cite{Bosch1992, Mavrin2017, Mavrin2018}. Throughout all the analyses in this paper we assumed the impurity density was a uniform fraction of the plasma density. This is a good approximation for L-modes, which have been shown to have a large amount of turbulent particle transport \cite{Rice2015}, and in accordance with the EU-DEMO design guidelines as well as previous reactor design studies \cite{Zohm2013,Sorbom2015,Buttery2021}. In future work we would like to include impurity profiles based on impurity transport calculations. Based on the resultant radiation from these calculations the impurity density was varied such that the tokamak power balance (\ref{eq:powbal}) produced $P_\mathrm{SOL} = 50$ MW $ \ll P_{LH}$, or  $f_{rad} \sim 85\%$ ($P_{LH}$ at the operating point we will show here was $\sim 130$ MW). Here we used a Kr radiator that radiates primarily in the plasma edge through line radiation rather than in the core by bremsstrahlung and synchrotron radiation avoid radiative collapse. Kr was the preferred radiator throughout this study rather than Xe as Kr produces a narrower radiation belt further off axis, reducing radiation at the $q=2$ surface where a locally negative power balance from radiation could serve as a trigger for tearing modes. This property of Kr is desirable enough to offset the small loss in fusion power that comes from the increased plasma dilution versus Xe.

We fixed minor radius $a = 1.2$ m; $H_{89} = 1.3$ (a conservative value for a radiatively improved L-mode based on review of the literature discussed in Sec.~\ref{sec:background}); $\kappa=1.6$ (well within empirical vertical stability limits \cite{Sweeney2020}); and the on-axis magnetic field $B_0 = 11.5$ T (similar to the ARC V0 design from Commonwealth Fusion Systems \cite{CreelyAPS2021}). The operating point's fusion power was determined by imposing a $\sim1$ MW/m$^2$ administrative limit on radiative wall loading, assuming a conformal vacuum vessel with surface area $S \sim S_{plasma}$.

The final operating point was found by minimizing the plasma current $I_p$ and maximizing $R$ while maintaining $q_* \ge 2.5$. Maximizing $R$ is desirable as it increases the space that the solenoid may occupy, increasing available flux and pulse time. Assuming that the blanket, shielding, and TF size remain roughly constant, available solenoid flux $\propto R^2$. Thus, increasing $R$ has a large effect on pulse length as $\tau_{pulse}$ scales linearly with the amount of available flux. A maximum $R$ for a fixed $q^*$ is obtained when the reduction in $n_g$ from dropping $I_p$ leaves a limited operating region at excessively high values of $n_e/n_g$ or closes the operating window completely. Considering these constraints values of $R = 4.2$ m and $I_p = 14$ MA were chosen for the remainder of this study. A complete list of the 0-D ARC RPL-mode scenario parameters can be found in Tbl.~\ref{tbl:0dparams}.

The POPCONs found in Fig.~\ref{fig:popcon} result from calculating the power balance based on the parameters in Tbl.~\ref{tbl:0dparams}. One surprising result of this analysis is that, despite the only slightly enhanced L-mode confinement, the plasma has a large ignition region and a thermally stable operating point is achievable. This results from the high operational densities enabled by the combination of high-field, compact size, and pulsed operation. Another attractive feature of the RPL-mode operation is an exceptionally small, $< 5$ MW, Cordey pass \cite{Cordey1987} indicating ARC tokamaks in RPL-mode could use rudimentary RF systems. However, our analysis here utilizes a larger more complex system in the interest of enhanced control. 

It is noted that the 0-D operating point chosen here is not absolutely optimized for pulse time. Greater $\tau_{pulse}$ is attainable, for example, if $R$ is increased to $\sim 4.5$ m and $I_p$ is decreased to $13$ MA (assuming solenoid flux which goes $\propto R^2$). However, absolute maximization in the 0-D model lead to operating points that were not viable when performance degradation from more sophisticated physics models was introduced later. Furthermore, we wanted to limit the major radius, making it no more than 1 m larger than the original ARC design, to maintain a compact form factor. 

\section{Integrated Modeling of an RPL-Mode Scenario}
\label{sec:integrated}

\begin{figure*}
    \centering
    \includegraphics[width=0.75\linewidth]{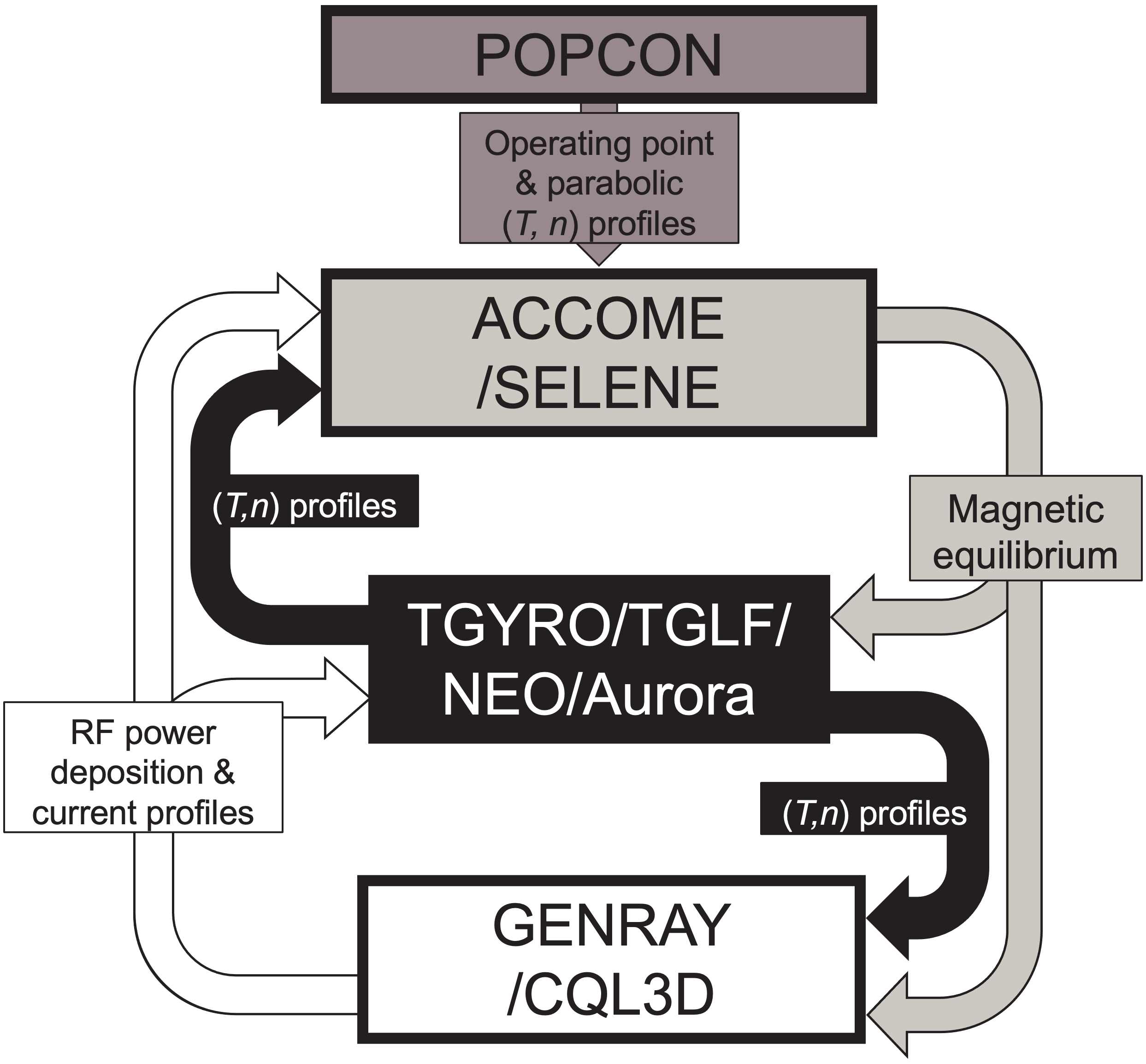}
    \caption{A diagram of the integrated RPL-mode model iteration loop.}
    \label{fig:iterationloop}
\end{figure*}

With the exception of TFTR, radiative L-modes haven't been demonstrated anywhere near fusion relevant conditions \cite{Hill1999NF,Hill1999PoP}. More rigorous analysis than simple application of scaling laws is necessary. Because of this we must supplement the scaling law analysis in Sec.~\ref{sec:0D} with a more sophisticated, self-consistent, integrated model. Our integrated model of a RPL-Mode scenario, shown schematically in Fig.~\ref{fig:iterationloop}, consists of three coupled simulation models: a free-boundary Grad-Shafranov (GS) simulation (ACCOME) \cite{Devoto1992,JAERI8042}, a 1D transport simulation (TGYRO/TGLF/NEO/Aurora) \cite{Belli2008,Belli2011,Candy2009,Candy2013,Staebler2007,Staebler2010,Staebler2021,Kernbichler2008,Sciortino2021}, and an LHCD simulation (GENRAY/CQL3D) \cite{Smirnov1994,Kerbel1985,CQL3D}. The integrated model was initalized by creating a magnetic equilibrium in ACCOME using the 0-D POPCON operating point parameters. ACCOME was chosen as our free boundary solver as it is computationally efficient and has a self-consistent loop-voltage model for ohmic plasmas. ACCOME generates a magnetic equilibrium by creating a separatrix boundary from shaping inputs, then iteratively solving the GS equation varying applied loop voltage and field coil currents while also accounting for the plasma bootstrap current and RF current drive. ACCOME performs a $\chi$-squared minimization between the GS solution obtained at each step in its iteration and the target separatrix boundary to determine the loop-voltage and coil currents needed to obtain the desired equilibrium shape. While ACCOME is capable of rudimentary loop voltage calculations, it does not include a sawtooth model. This means the current profiles produced are overly peaked as they have not been relaxed by internal MHD modes. The lack of sawtooth model is a known deficiency of the simulation workflow here and future work will attempt to integrate a more sophisticated magnetic equilibrium calculation with a loop-voltage/sawtooth model such as TSC \cite{Jardin1986,Jardin1993}.

\begin{figure}
    \includegraphics[width=1\linewidth]{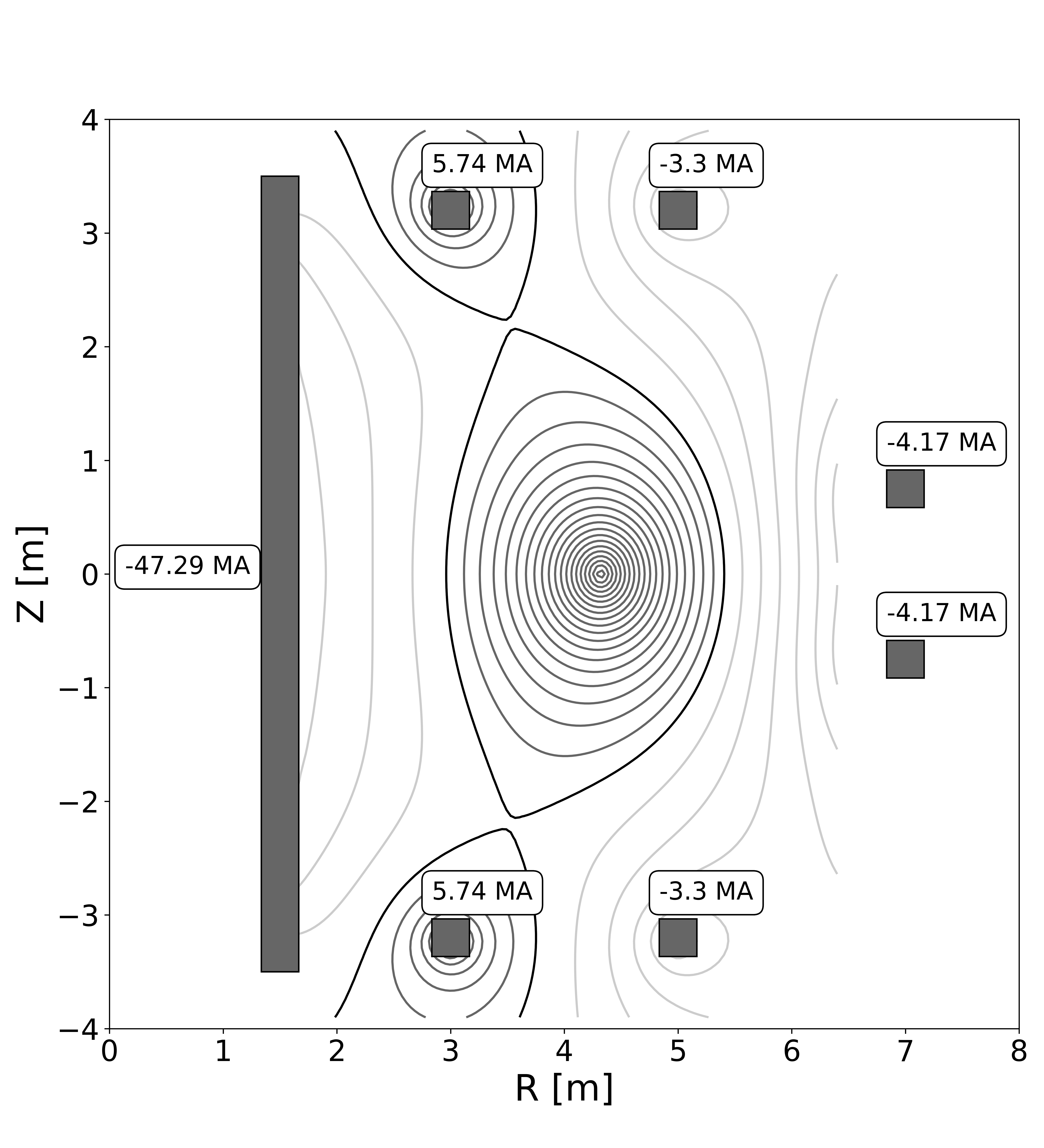}
	\centering
	\caption{An RPL-mode free boundary equilibrium solution generated with ACCOME.}
	\label{fig:equ}
\end{figure}

\begin{figure}
	\includegraphics[width=0.85\linewidth]{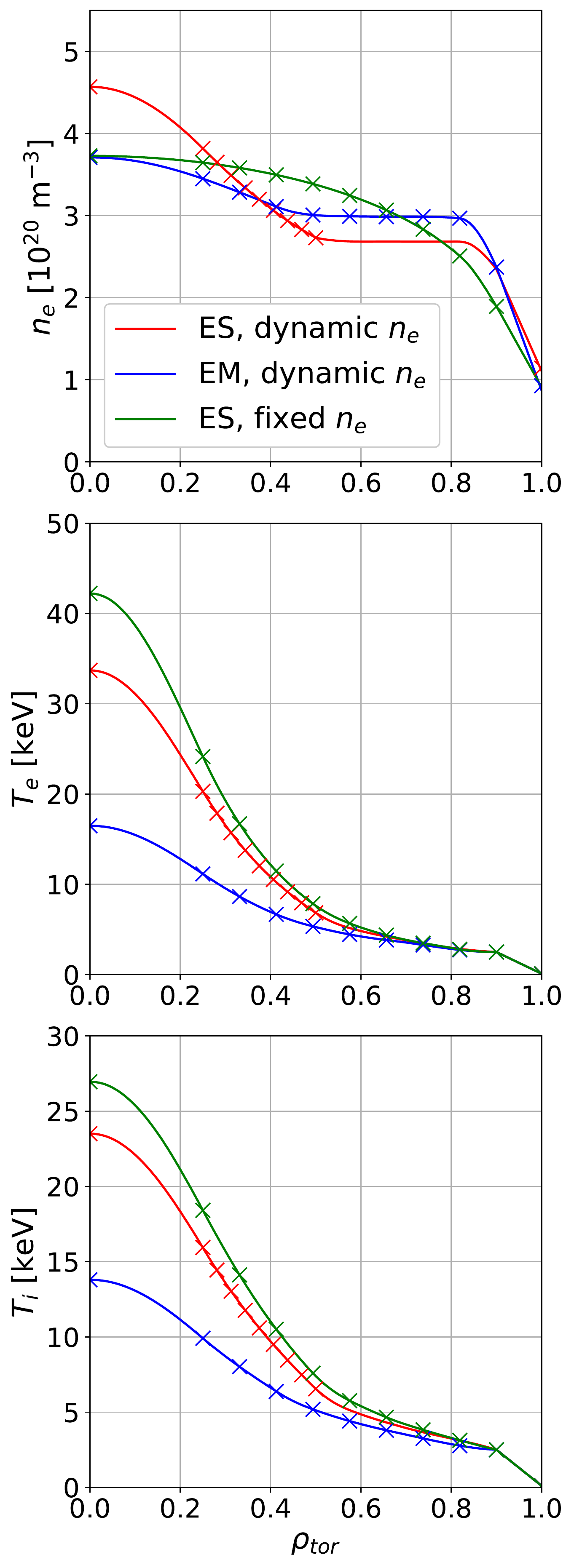}
	\centering
	\caption{Radial (\textit{T,n}) profiles obtained from transport simulations using electrostatic (ES) and electromagnetic (EM) SAT2 with both dynamically calculated density profiles and fixed profiles using the Angioni scaling (\ref{eq:angioni}) \cite{Angioni2007,Angioni2009}.}
	\label{fig:prof1d}
\end{figure}

The ACCOME equilbrium solution and parabolic POPCON plasma profiles were used to initialize a TGYRO simulation with TGLF-SAT2 \cite{Staebler2021} for calcuation of 1-D turbulent transport to produce (\textit{T,n}) profiles. In the absence of a sawtooth model, to prevent magnetic shear from being over-predicted in TGYRO/TGLF simulations we have forced the safety factor $q \rightarrow 1$ within the region where $q<1$ in the equilibria. To demonstrate the RPL-mode's stability against radiative collapse impurity radiation was calculated self-consistently in our model. Radiation coefficients from Aurora were used to assess the radiation profiles and radiated power fraction and the computed radiation profiles were included in the TGYRO transport calculations. The (\textit{T,n}) profiles produced by the transport simulations were passed to ACCOME and the equilibrium was recalculated. Transport calculations were then rerun using the updated equilibrium. This process was iterated until the temperature profiles and equilibrium stopped evolving indicating convergence. 

\begin{figure}
	\includegraphics[width=1\linewidth]{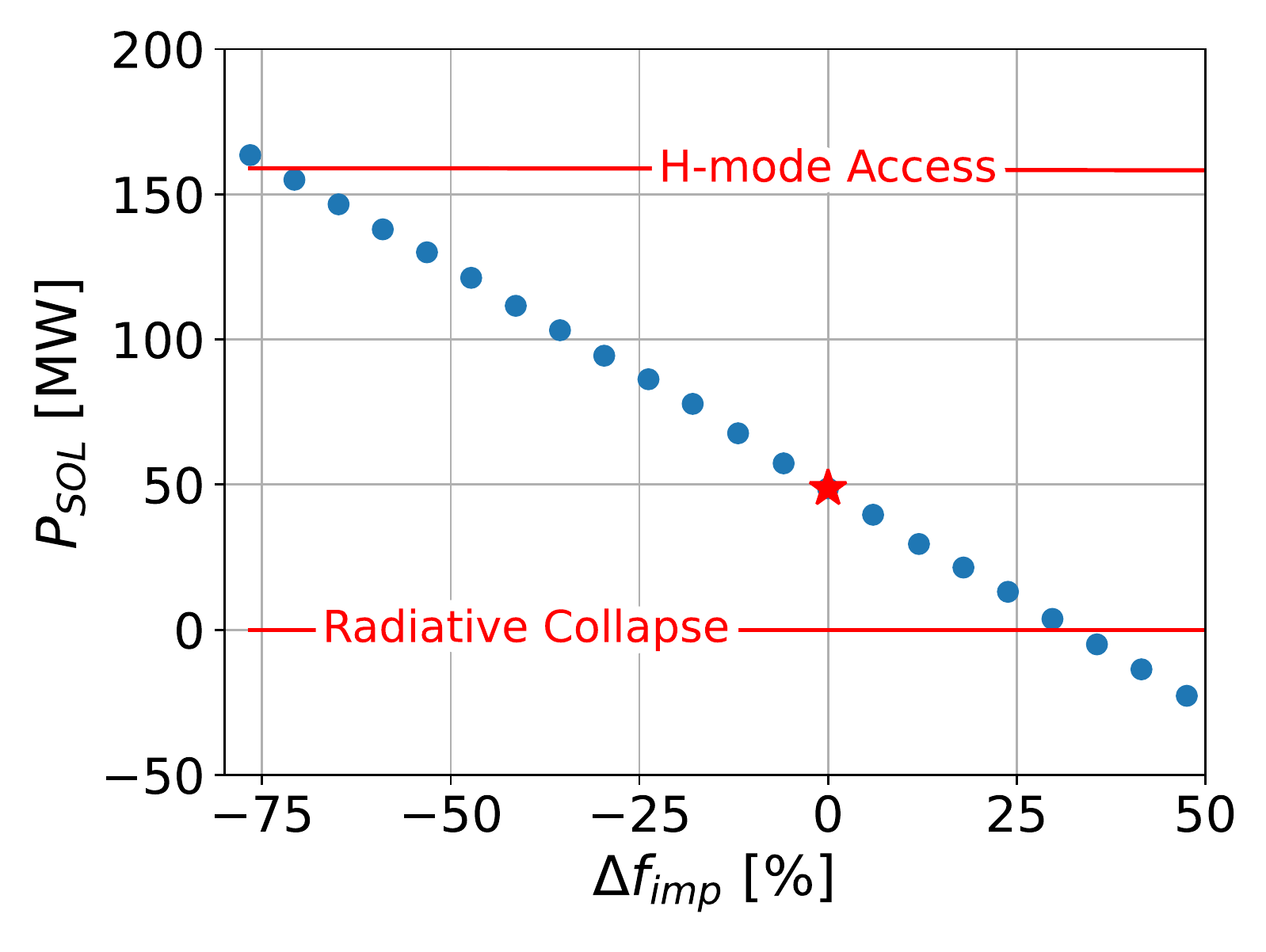}
	\caption{Percent variance in impurity fraction from the ES-SAT2 Dynamic-n baseline, $\Delta f_{imp}$, versus the total power transported to the scrapeoff layer $P_{SOL}$ (negative values indicate decreasing impurity fraction positive values indicate increasing impurity fraction). The red lines represent the L-mode access (upper) and radiative collapse (lower) limits. The red star represents our nominal operating point and blue dots are off-normal operating points at differing impurity fractions.}
	\label{fig:sensitivity}
\end{figure} 

Examples of an RPL-mode equilibrium and the plasma profiles generated by the integrated simulation model are shown in Figs.~\ref{fig:equ} and \ref{fig:prof1d}. Unlike H-mode simulations the peeling ballooning limit \cite{Snyder2011} cannot be used to set a boundary condition, making boundary conditions in L-mode difficult to determine without prior experimental motivation. The TGYRO transport simulations used boundary values of $T = 2.5$ keV and $n = 0.25\, n_0$ at $\rho = 0.9$ based on Alcator C-Mod L-modes \cite{Greenwald1997,Schmidt2011} (calculations of confinement enhancement based on the final TGYRO results indicated these boundary conditions accurately replicate L-mode confinement, $H89 \le 1.2$). TGYRO simulations using electrostatic (ES) and electromagnetic (EM) TGLF-SAT2 transport rules \cite{Staebler2021} were performed over 9 points equispaced from $\rho = 0.25$ to 0.9, where $\rho$ is the normalized square root of the toroidal flux. Transport simulations with computed and imposed density profiles based on empirical peaking factors were evaluated \cite{Angioni2007,Greenwald2007,Takenaga2008,Angioni2009}. Imposed density profiles used the Angioni scaling for density peaking:
\begin{equation}\label{eq:angioni}
    \nu_{ne}^{Angioni} = 1.347 - 0.117 \ln \nu_{eff} - 4.03 \beta 
\end{equation}
\begin{equation}
    \nu_{eff} = \frac{0.1 Z_{eff}\langle n_e \rangle R_0}{\langle T_e \rangle^2}
\end{equation}
\begin{equation}
    \beta = \frac{4.02 \times 10^{-3}\langle p \rangle}{B_0^2},
\end{equation}
where $\langle n_e \rangle$ is volume averaged electron density in $10^{19}$ m$^{-3}$, $\langle T_e \rangle$ is the volume averaged electron temperature in keV, and $\langle p \rangle$ is the volume averaged pressure in keV $10^{19}$ m$^{-3}$. The Angioni scaling is technically only valid for H-modes, but it has been used in a similar fashion for prediction of L-mode performance in SPARC \cite{Creely2020} as no L-mode scaling exists. Empirical peaking factors were studied because in high field devices TGYRO with TGLF-SAT2 has been shown to underpredict density peaking, producing unphysically flat profiles in high-field scenarios \cite{RF2022}. In fact, in ES-SAT2 even at zero density gradient, TGLF predicted a net particle flux for $\rho>0.6$, affecting convergence. As such, this case was iterated between predicted density/temperature profiles between $\rho = 0.25$ to 0.9 and $\rho = 0.25$ to 0.5 until profiles were flux-matched. This was not necessary for the EM-SAT2 case, but flux-matching still showed essentially zero density gradient for $\rho>0.6$ (this explains the difference between the ES-SAT2 tick locations and the other cases in Fig.~\ref{fig:prof1d}). To study the robustness of the operating point to variance in the impurity fraction, the Kr impurity fraction was scanned. The results of this scan performed in the ES dynamic-n TGYRO case shown in Fig.~\ref{fig:sensitivity} demonstrate that a greater than $70\%$ change in the impurity fraction is required to trigger entrance into H-mode, and a $\sim 25\%$ change in impurity fraction is necessary to induce radiative collapse. It must be noted that, for simplicity, these simulations did not include impurities outside the introduced Kr impurity. In a real device, intrinsic impurities from PFCs, such as C and W, would also contribute to the core radiation power balance, and the Kr impurity fraction varied accordingly to avoid radiative collapse. Intrinsic impurities would also increase fuel dilution $f_{dil}$, reducing fusion power for fixed $f_g$. Furthermore, it is uncertain whether or not impurity transport in these situations will be neoclassical or turbulent. Future gyrokinetic and neoclassic impurity transport simulations will need to be performed to determine the degree of impurity peaking that can be expected in these highly radiative ARC scenarios.

\begin{figure}
    \centering
    \includegraphics[width=0.95\linewidth]{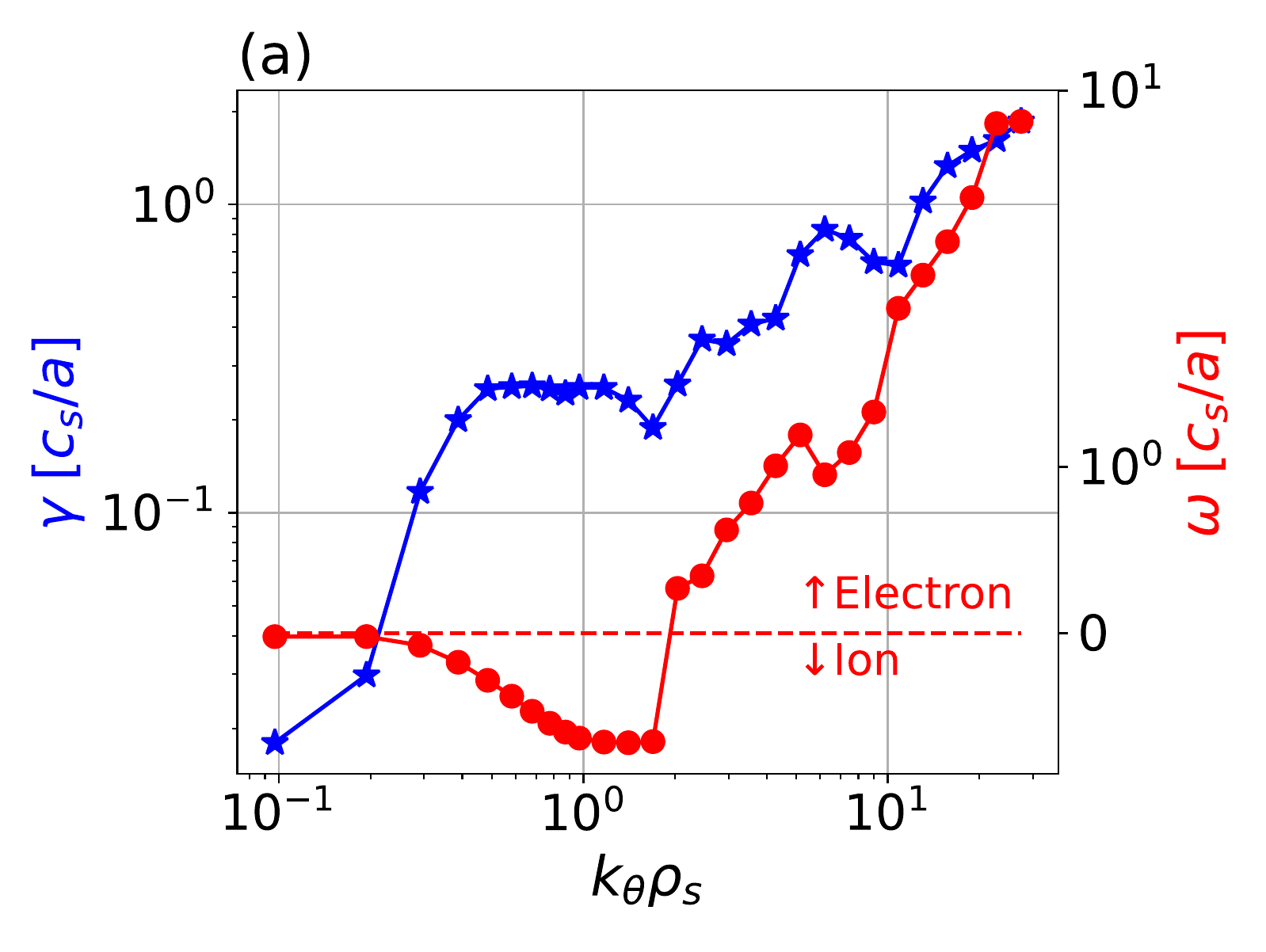}
    \includegraphics[width=0.9\linewidth]{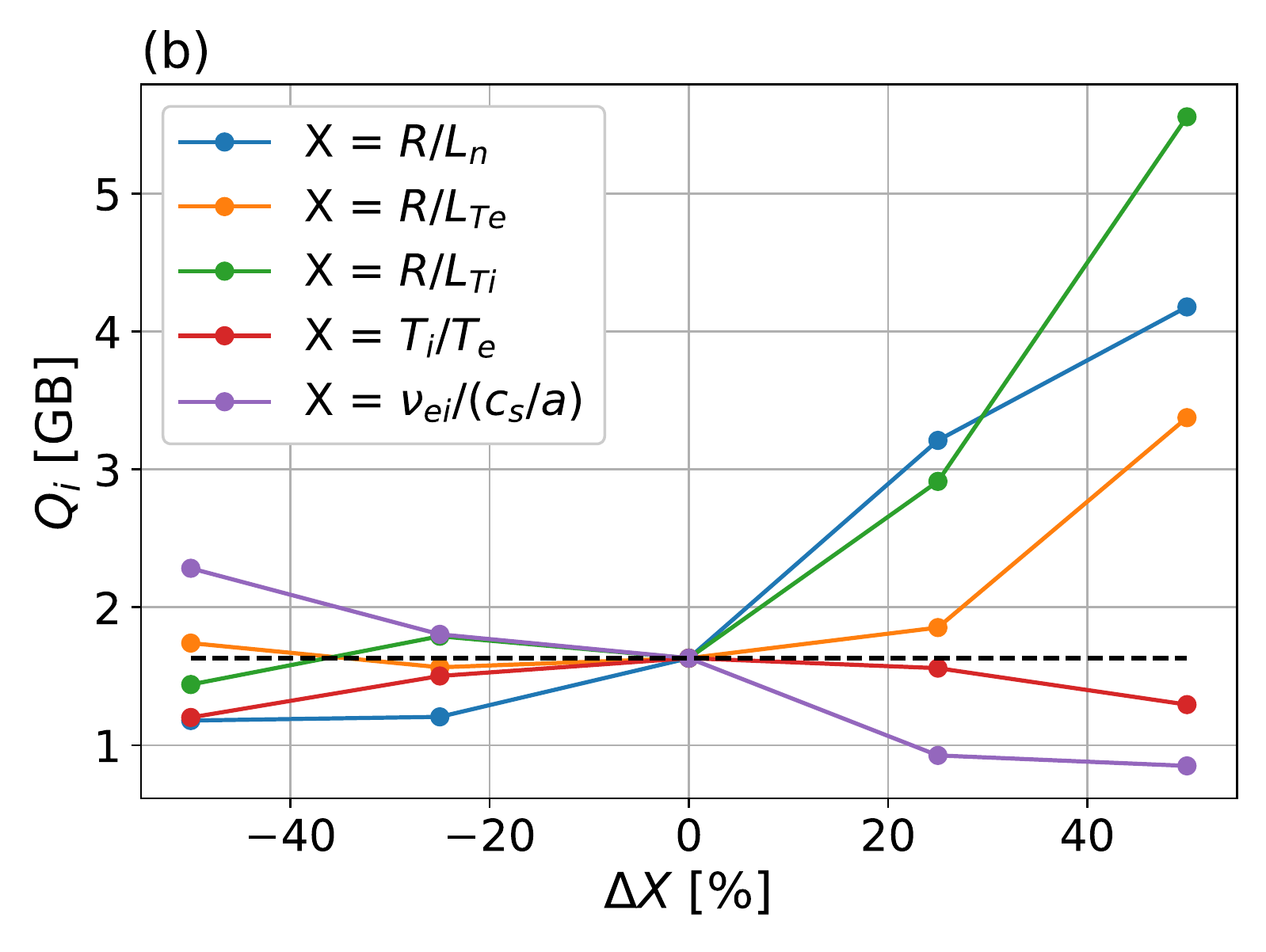}
    \caption{(a) TGLF turbulence linear frequency (dots) and growth rate (stars) spectra for ES-SAT2 with dynamic density profiles. (b) Variation in the TLGF ion heat flux at $\rho=0.35$ versus standard turbulence drives: normalized electron density gradient $R/L_n$, electron temperature gradient $R/L_{Te}$, ion temperature gradient $R/L_{Ti}$, ion-electron temperature ratio $T_i/T_e$, and normalized electron-ion collision frequency $\nu_{ei}/(c_s/a)$ where $c_s$ is the ion sound speed.}
    \label{fig:tgyro_spectrum}
\end{figure} 

Analysis of the turbulent spectrum produced by TGYRO/TGLF and the ion heat fluxes' dependence on turbulent drives, shown in Fig.~\ref{fig:tgyro_spectrum}, indicates that transport is ITG dominated. This is the expected result for reactor relevant regimes where normalized gyroradius $\rho^* = \rho_i / a$ is small and the collisionality normalized by the bounce frequency $\omega_{B,e}$, $\nu^* \sim \nu_e / \epsilon \omega_{B,e}$, is large. We also attempted to replicate the turbulence suppression through impurity injection observed in the radiative L-mode experiments discussed in Sec.~\ref{sec:background} by varying the impurity fraction and evaluating the ITG growth rate normalized against the local gradients. However, ITG suppression was only observed using SAT1 which consistently demonstrated ITG suppression as a result of enhanced $Z_{eff}$ from impurity seeding, but also produced unrealistic temperature profiles and was generally unstable with poor numerical convergence. The stiffer TGLF-SAT2 model used to produce our results demonstrated better numerical convergence but did not exhibit changes in the normalized ITG growth rate or core peaking with varying impurity fraction. 

The electromagnetic version of TGLF-SAT2 greatly decreased the plasma performance. This was unexpected as $\beta_N$ values in RPL-modes are low relative to other reactor relevant scenarios. The final turbulent growth rates and spectra in the EM and ES cases were similar, however, the critical gradients in the EM case were much lower limiting on axis temperature and density peaking substantially. Reduced density peaking in reactor relevant scenarios has been observed in previous comparisons of EM and ES TGLF, but the temperature peaking reduction was much less severe \cite{Fable2019}. As RPL-modes are at substantially lower $\beta_N$ than previous reactor relevant plasmas which were analyzed, EM effects should be of less importance, and it is possible that the performance reduction observed here may be an artifact of the RPL-modes being outside the range of validity of EM TGLF-SAT2. Full gyrokinetics simulations may be necessary to determine the correct behavior when EM effects are included. 

Using the electrostatic SAT-2 operating point in Tbl.~\ref{tbl:intresults}, it is possible to estimate the parallel heat flux that will need to be mitigated by the divertor in an ARC RPL-mode. Using the Brunner scaling for scrape-off-layer heat flux width \cite{Brunner2018}:
\begin{equation}
    \lambda_q \, [\textrm{mm}]= 0.9 (\langle p \rangle [\textrm{atm}])^{-0.48},
\end{equation}
where $\langle p \rangle$ is the volume averaged plasma pressure, yields heat flux widths in our RPL-mode ARC of 0.37 mm and the pulsed inductive ITER baseline of 0.57 mm \cite{Shimada2007}. It is notable that the Brunner scaling is the most conservative scaling for scrape-off-layer heat flux width and generally predicts smaller $\lambda_q$ values than other scaling laws such as the Eich scaling (regression \#14) \cite{Eich2013}. However, the Eich scaling is only strictly valid for H-mode plasmas while the Brunner scaling is intended to be operation mode agnostic \cite{Brunner2018} (and predicts smaller L-mode $\lambda_q$ values in high $\langle p \rangle$ plasmas than L-mode specific scaling laws which, for example, predict $\lambda_q \sim 5$ mm in an RPL-mode ARC \cite{Horacek2020}). We may estimate the parallel heat flux to be \cite{Stangeby}:
\begin{eqnarray}
    q_\parallel &= f_{target} \frac{P_{SOL}}{2\pi (a+R) \lambda_q} \frac{B_\phi}{B_\theta} \\ &\simeq f_{target} \frac{P_{SOL} q_{95}}{2\pi \epsilon (a+R)\lambda_q} ,    
\end{eqnarray}
where $f_{target}$ is the fraction of the total parallel heat flux directed at a given divertor target, and $q_{95}$ is the safety factor at the normalized square-root toroidal flux $\sqrt{\phi_n} = 0.95$ surface. Assuming $f_{target} = 0.35$ for the outer target of the double null divertor in ARC and $f_{target} = 0.5$ in the ITER single-null divertor (equal power sharing) \cite{Brunner2018_2}, we may estimate the $q_\parallel$ in each case. This analysis estimates that the ARC and ITER divertors will need to mitigate $q_\parallel$ of 15.4 GW/m$^2$ and 16.1 GW/m$^2$ respectively (the less pessimistic Eich reg. 14 scaling, assuming $B_\theta \sim a B_0 / R_0 q_{95}$, predicts $q_\parallel$ of 9.45 GW/m$^2$ and 7.05 GW/m$^2$ in ARC and ITER respectively). This analysis suggests that the divertor challenge in an RPL-mode ARC reactor is roughly similar to that in ITER, despite the fusion power density being $\mathcal{O}(10)$ times larger, and it suggests a conventional divertor configuration could be used in an RPL-mode ARC unlike previous ARC designs \cite{Kuang2018,Wigram2019}.

\begin{table*} 
\centering
\caption{\label{tbl:intresults} Integrated simulations results for ARC RPL-Mode scenarios with various SAT rules. A dash `-' indicates a parameter that was fixed for all simulations.}
\begin{tabular}{lrrrr}
\hline
Parameter & $\quad$ & ES fixed n & ES SAT2 dyn. n  & EM SAT2 dyn. n\\
\hline
R & [m] & 4.18 & 4.18 & 4.19  \\
a & [m] & 1.21  & 1.22 & 1.21 \\
$I_p$ & [MA] & 14 & - & -  \\
$B_0$ & [T] & 11.5 & - & - \\
$\kappa_{95}$& & 1.54 & 1.54 & 1.62  \\
$\delta_{95}$& & 0.23 & 0.25 & 0.35 \\
$T_{e,0}$& [keV] & 42.2 & 33.7 & 16.5\\
$T_{i,0}$& [keV] & 27.0 & 23.5 & 13.8\\
$n_0$& $[10^{20}$ m$^{-3}]$ & 3.73 & 4.57 & 3.71\\ 
$f_g$ & &  0.91 & 0.91 & 0.92\\
$\beta_N$& & 1.4 & 1.2 & 0.8 \\
$f_{imp}$& & 4.9e-4 & 4.2e-4 & 6.0e-5\\
$f_{dil}$& & 0.94 & 0.95 & 0.96 \\
$q_{95}$& & 2.9 & 3.1 & 3.2 \\
$f_{bootstrap}$& & 0.18 & 0.16 & 0.12\\
$V_{loop}$& [V] & 0.135 & 0.157 & 0.365 \\
$P_{fus}$& [MW] & 1168 & 1042 & 334 \\
$P_{SOL}$& [MW] & 53.0 & 47.1 & 46.1 \\
$f_{rad}$& & 0.80 & 0.80 & 0.52\\
$P_{LHCD}$& [MW] & 24.0 & 23.5 & 22.7  \\
$f_{LHCD}$& [GHz] & 8 & - & - \\
$n_{\parallel,LHCD}$& & 2.1 & 2.2 & 2.1 \\
$\theta_{launcher}$& [deg] & 80 & 75 & 80 \\
$\eta_{20,CD}$& & 0.084 & 0.082 & 0.080\\
\hline
\end{tabular}
\end{table*}

\begin{figure}\label{fig:rf}
	\includegraphics[width=\linewidth]{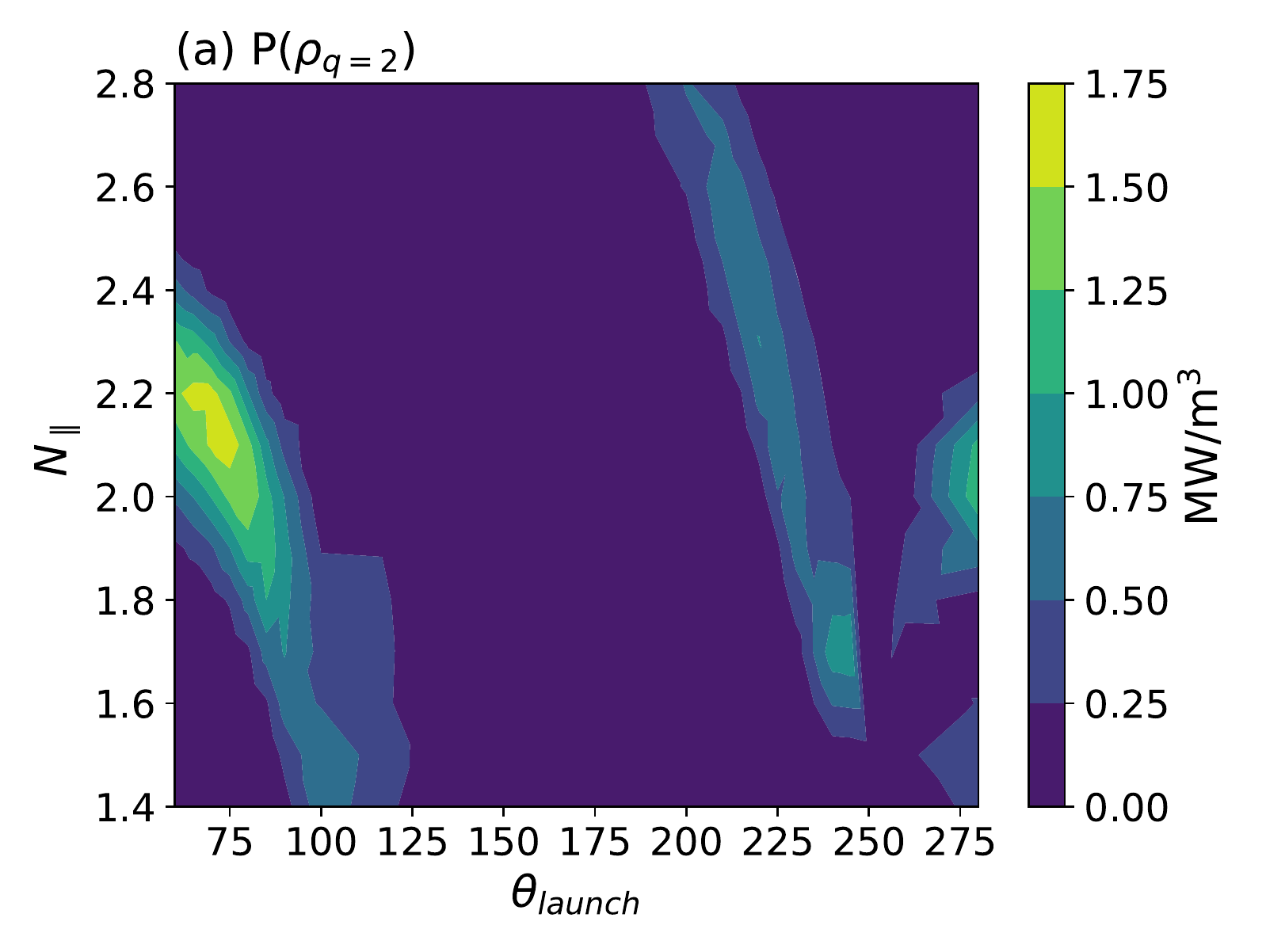}
	\includegraphics[width=\linewidth]{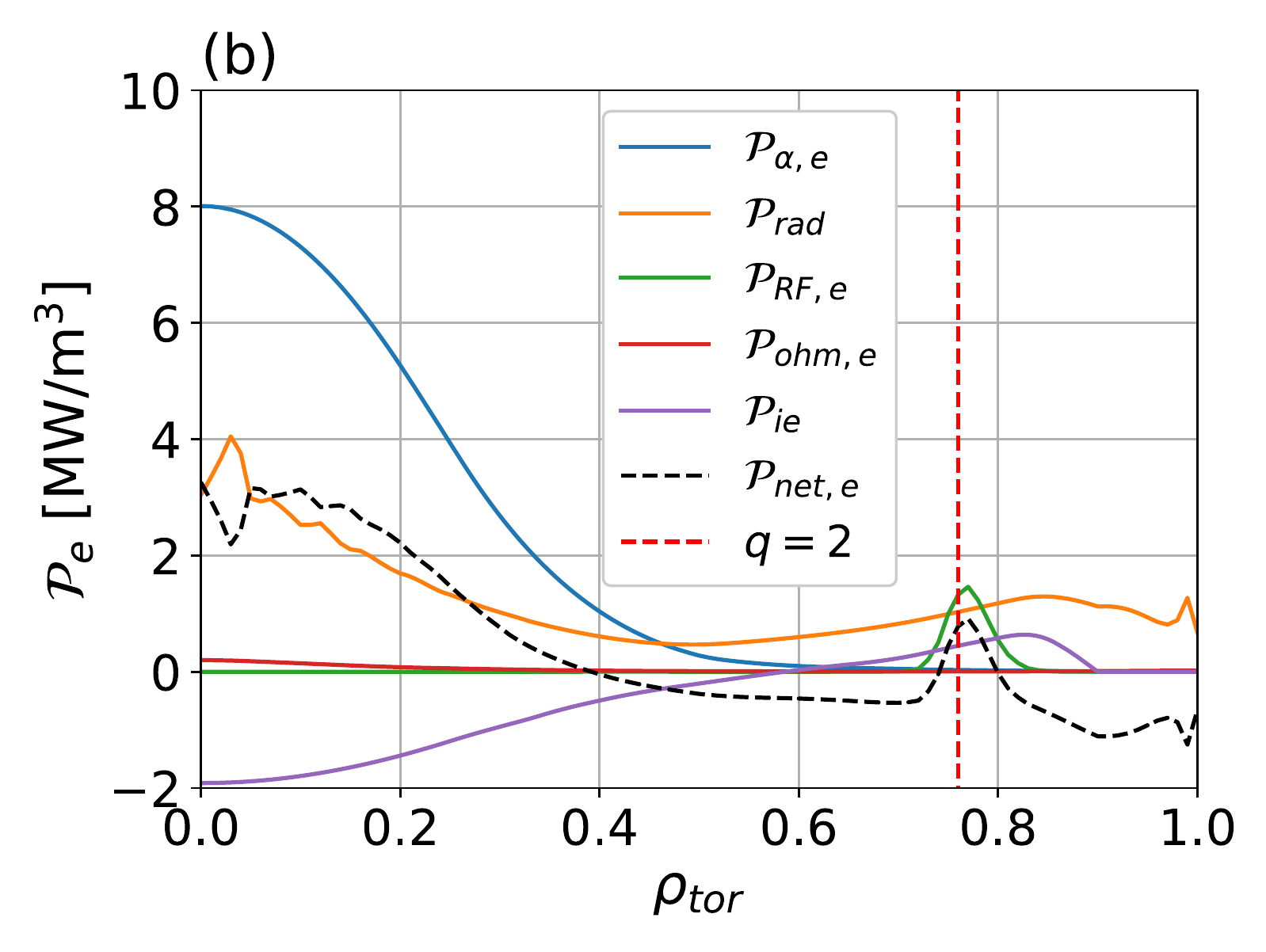}
	\caption{(a) Contours showing power on the $q=2$ flux surface versus $n_\parallel$ and $\theta_{launch}$ for the ES-SAT2 case with dynamic density profiles. (b) Power density profiles for the electron channel using the ES-SAT2 case with dynamic density profiles and a LHCD launcher with $n_\parallel=2.1$ and $\theta_{launch} = 75$ degrees.}
\end{figure}

Initial iterations of the integrated model used a simple 20MW Gaussian electron heating source, driving 0.5 MA of current, placed off axis about the $q=2$ surface. After the model began to converge with the simplified heating source, the RF heating and current drive profiles calculated by GENRAY/CQL3D were used instead. LHCD access is described by accessibility conditions \cite{Bonoli1985}: 
\begin{equation}\label{eq:lhaccess}
    \frac{\omega_{pe}}{\Omega_e} + \sqrt{1+\frac{\omega_{pe}^2}{\Omega_e^2}-\frac{\omega_{pi}^2}{\omega^2}} \leq n_\parallel < \frac{c}{3v_{the}},
\end{equation}
where $\omega_{ps} = \sqrt{n_s e^2 / \epsilon_0 m_s}$ is the plasma frequency for species $s$, $\Omega_e = eB/m_e$ is the electron cyclotron frequency and $v_{the} = \sqrt{2 T_e / m_e}$ is the electron thermal speed. This prescribes that $n_{\parallel,max} \propto 1/\sqrt{T_e}$ and $n_{\parallel, min} \propto \sqrt{n_e}/B$. The high-field and the L-mode profiles which do not have a large temperature and density pedestal broaden the access window. This allowed us to design a current drive actuator for our RPL-mode scenario capable of delivering power to the $q=2$ surface without encountering a cutoff. Raytracing/Fokker-Planck simulations like GENRAY/CQL3D should be accurate in ARC scenarios as turbulence and diffraction are of little importance when the LHCD is single-pass damped \cite{Biswas2020,Frank2022}. There should also be minimal power loss to parametric decay as the LHCD here has $\omega > 2\omega_{LH}$ in the edge, where $\omega_{LH}$ is the lower-hybrid frequency \cite{Porkolab1977,Cesario2004}. The use of the higher 8 GHz source frequency in this ARC scenario relative to the 4.6 GHz source frequency found in most experiments should help to minimize parametric decay and the deleterious effects of the LHCD density limit \cite{Wallace2011,Baek2018,Ding2018}. Higher source frequencies have been demonstrated to increase the LHCD density limit in experiments in FTU \cite{Pericoli1999,Cesario2004}.

The tearing mode stabilization scheme utilized in this RPL-mode ARC scenario is more sophisticated than that found in previous experimental investigations of LHCD suppression of tearing modes \cite{Warrick2000}. The current drive and heating in this scenario \textit{does not} have a broad deposition profile and is instead highly localized on the $q=2$ flux surface. Through use of the non-linear RF condensation effect, where the temperature perturbation resultant from magnetic island formation due to tearing may be used to further localize the RF-current drive \cite{Reiman2018,Frank2020}, island stabilization can be obtained at much lower RF powers and total driven currents. Since the current drive is targeted, only the local bootstrap current density near the $q=2$ flux surface must be replaced by LHCD rather than a large fraction of the entire bootstrap current \cite{Jin2021}. In order to determine the optimal launch parameters to deliver power to the $q=2$ flux surface for tearing mode suppression, we performed a parametric scan of both the launched parallel refractive index $n_\parallel$, and the poloidal launch location $\theta_{launch}$. GENRAY/CQL3D simulations were performed using 264 different launcher configuration with $n_\parallel = 1.5$ to 2.6 and $\theta_{launch} = 70$ to 280 degrees from the outboard mid-plane. Additional grid refinement simulations were then performed after the initial scan around the locations of peak power deposition. The launcher configuration which delivered the greatest power density to the $q=2$ surface was then chosen. Raytracing simulations used 75 rays, a spectral width of $\Delta n_\parallel = 0.4$, and a launcher height of $0.75$ m. The results of this scan and the ARC RPL-mode power density profiles produced with the final current drive system in the ES-SAT2 case with dynamic density profiles are shown in Fig.~\ref{fig:rf}. All GENRAY/CQL3D simulations used 25MW of LHCD power, but small adjustments to the LHCD power were made in the transport calculations to fine tune the power balance on the $q=2$ surface. Such modifications are acceptable as we found power deposition location was insensitive to small changes in launched power. Current drive efficiency in these simulations was generally low, $\eta_{20,CD} = 0.082$ in the ES-SAT2 case with dynamic density profiles where 175 kA of total current were driven (the LHCD is linearly damped on the relatively cold, $\sim 5$ keV, $q=2$ surface so $n_\parallel$ is large lowering $\eta_{20,CD}$). However, the local current density at the $q=2$ surface from LHCD $J_{LH} \sim$200 kA/m$^2$, was large relative to the local current density from the bootstrap current of $J_{BS} \sim$125 kA/m$^2$. 

To confirm the LHCD systems here are actually effective at suppressing tearing modes, two fluid modeling of magnetic island auto-stabilization, using the method reported in \cite{Jin2021,Jin2021b} was performed. These calculations used the following parameters: a local LH driven electron power density of 1 MW/m$^3$ based on the integrated modeling results shown in Fig.~\ref{fig:rf}, magnetic island electron and ion temperatures of 3 keV and 4 keV respectively based on the TGYRO results shown in Fig.~\ref{fig:prof1d}, island electron densities of $3\times10^{20}$ m$^{-3}$ also based on the TGYRO modeling results, and a conservative LH to bootstrap current ratio of 0.5 to 1.0 based on the relative currents in the ACCOME and CQL3D modeling results. Perpendicular electron thermal diffusivity in the magnetic island was assumed to be $\sim 0.1$ m$^2$/s (this is an order of magnitude estimate and could be refined in future work). For these parameters, it was predicted that islands would be automatically stabilized for island widths $\sim$ 4 to 5 cm. These results indicate the LHCD systems here will automatically stabilize tearing modes at inconsequentially small island widths via RF condensation.

A summary of the results of our integrated modeling is in Tbl.~\ref{tbl:intresults}. These results demonstrate that viable RPL-mode operation points can be obtained using integrated simulations, and they are not simply an artifact of scoping using POPCONs with empirical confinement scalings and parabolic profiles. Despite L-mode confinement and high $f_{rad}$, the high $P_{fus}$ obtained in RPL-modes from density maximization maintains the power balance and avoids radiative collapse. 

\section{Extrapolation to Negative Triangularity}\label{sec:negD}
\begin{figure}
	\includegraphics[width=0.9\linewidth]{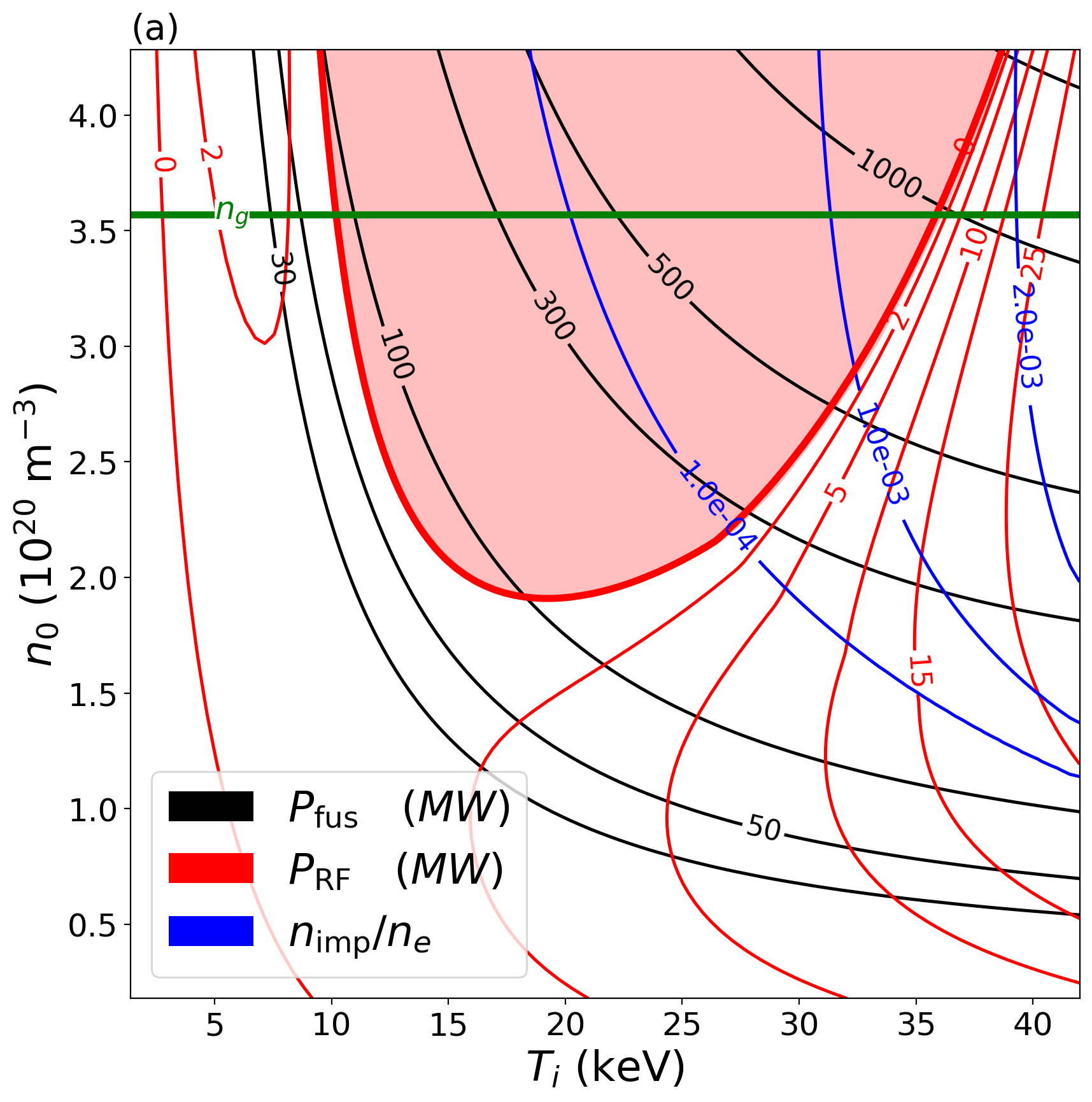}
	\includegraphics[width=1\linewidth]{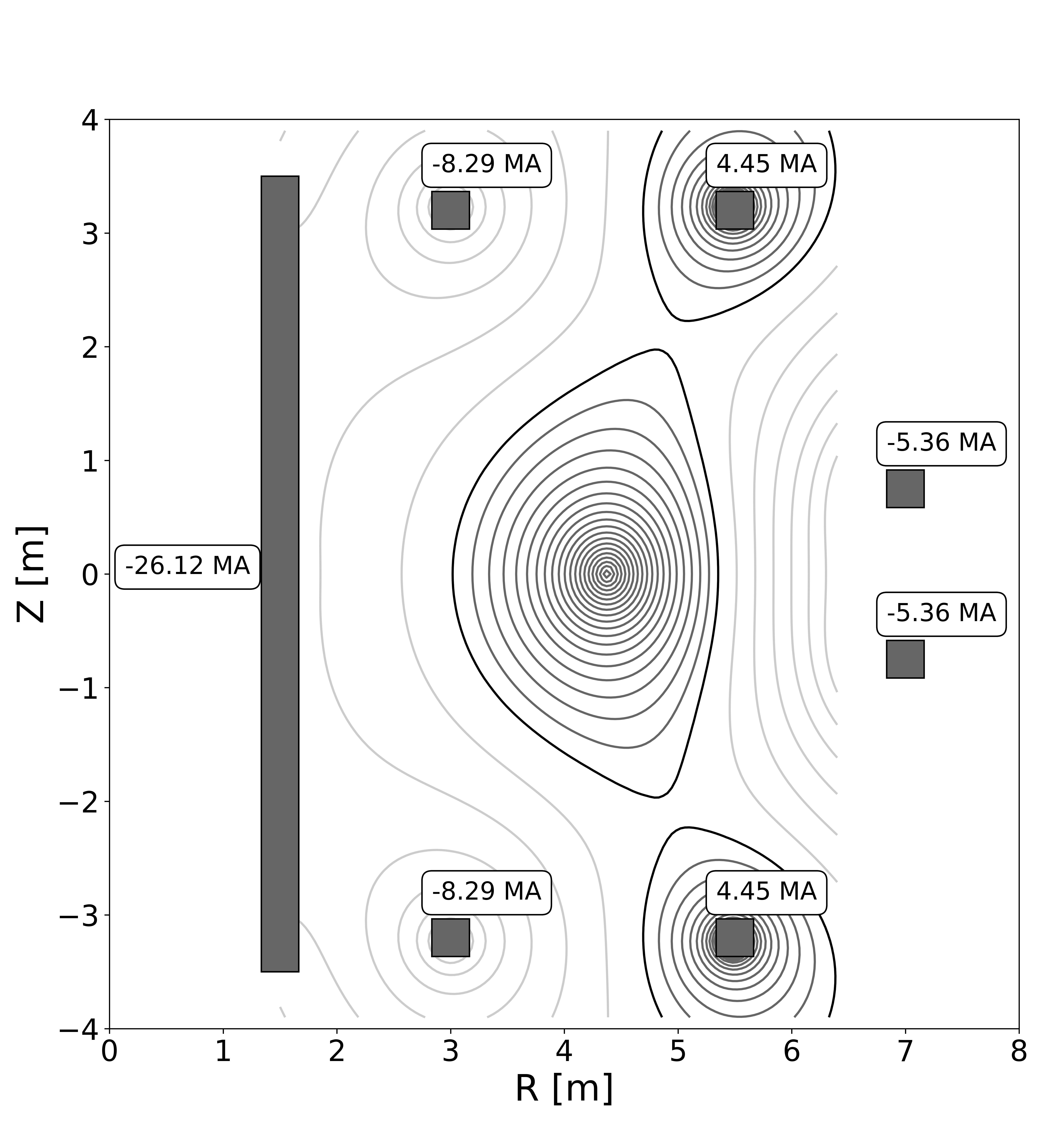}
	\caption{(a) Negative Triangularity POPCONs and (b) GS equilibrium}
	\label{fig:negD}
\end{figure}
Negative triangularity (NT) provides an attractive avenue to augment the performance of RPL-modes. NT has been investigated as a means to suppress trapped electron mode turbulence dominated in TCV \cite{Moret1997, Pochelon1999,Marinoni2009,Fontana2017,Huang2018}, however, recent modeling using nonlinear gyrokinetics \cite{Kinsey2007,Highcock2018,Duff2022} and further experimental results on TCV and DIII-D \cite{Marinoni2019,Fontana2019,Austin2019,Coda2021} have shown that NT may also suppress ITG turbulence in reactor-relevant regimes. Furthermore, diverted NT discharges \cite{Marinoni2021,Coda2021} with low field strike points at large major radius are anticipated to have easier divertor engineering \cite{Medvedev2015,Kikuchi2019,Marinoni2021}. NT also helps preserve L-mode by greatly increasing the H-mode power threshold, sometimes even eliminating the L-H transition entirely \cite{Austin2019,Marinoni2019,Marinoni2021,Coda2021,Nelson2022}. This prevents the situation in which impurity feed fails, leading to a transition to H-mode where enormous transient fusion powers and heat loads would likely cause severe damage to reactor components. Finally, enhanced confinement from NT enables reduced $I_p$ operation while maintaining large fusion power densities. Reduced $I_p$ operation in NT is extremely attractive as it will make disruptions less violent and reduce the loop voltage required to sustain the discharge.

\begin{table*}
\caption{\label{tbl:negD} 0-D Parameters of ARC Negative Triangularity Scenario}
\begin{tabular}{lr}
\hline
Parameter\\
\hline
Major Radius (R) & 4.18 m \\
Minor Radius (a) & 1.16 m \\
Plasma Current $\left(I_p\right)$ & 12 MA \\
Magnetic Field $\left(B_0\right)$ & 11.5 T \\
Elongation $\left(\kappa\right)$ & 1.6 \\
Triangularity $\left(\delta\right)$ & -0.5 \\
Confinement Enhancement Factor $\left(H_{89}\right)$ & 1.8 \\
Edge Safety Factor $\left(q_{95}\right)$ & $3.0$ \\
Greenwald Fraction $\left(f_g\right)$ & $\sim 1.0$ \\
Normalized Beta $\beta_N$ & 1.6 \\
RF Power $\left(P_{RF} \right)$ & 25 MW \\
Fusion Power $\left(P_{fus} \right)$ & 1150 MW \\
Power Conducted to SOL $\left(P_{SOL}\right)$ & 50 MW \\
$f_{bootstrap}$ & 0.25\\
Loop Voltage $\left( V_{loop} \right)$ & $6.7 \times 10^{-2}$ V \\
\hline
\end{tabular}
\end{table*}

We performed an enhanced POPCON scoping of NT RPL-modes in an ARC-class device. POPCONs were preferred over TGYRO using TGLF-SAT2 in this case as the validity of TGLF in negative triangularity scenarios still requires experimental validation. To perform our analysis, we generated a free boundary GS ARC equilibrium with negative triangularity, $\delta_{95} \sim -0.5$, using the pressure profiles from the integrated positive triangularity L-mode simulation using ES-SAT2 TGLF. The free boundary GS solution equilibrium geometry and the (\textit{T,n}) profile form factors obtained from TGYRO with ES-SAT2-TGLF were used to perform a `physics-informed' POPCON analysis similar to that found in \cite{Rodriguez-Fernandez2020} (rather than simple parabolic plasma profiles with equal ion and electron temperatures found in Sec.~\ref{sec:0D}). Assuming NT provides enhanced confinement similar to experimental discharges in this case, we fixed $H_{89} = 1.8$. We then iterated the physics informed POPCON with the GS solution until we found the minimum plasma current at which we could still produce a $P_{fus} \sim 1200$ MW operating point using the ES-SAT2-TGLF (\textit{T,n}) profiles.

The plasma current in our negative triangularity scenario could be reduced $\sim 15 \%$ to $I_p = 12$ MA relative to its positive triangularity counterpart. Further reduction in current was prevented by $n_g$ decreasing. Any further drop in current lead to substantial loss of fusion power relative to the positive triangulariy ES-SAT2 baseline. However, because of its enhanced confinement, the negative triangularity plasma had increased core temperatures and relative bootstrap fractions leading to a profound (factor of $\sim 2$) drop in the plasma loop voltage from the positive triangularity baseline (Our NT equilibrium simulations used a Gaussian 250kA, 25MW, current drive and heating source centered at the $q=2$ surface. This current drive source had the same CD efficiency as the converged positive triangularity baseline). This agrees with other recent negative triangularity scoping studies which also observed noticeably enhanced bootstrap fractions \cite{Schwartz2022}. Lower $V_{loop}$ will enable longer pulses in negative versus positive triangularity discharges. Longer pulses are beneficial to reactor economics as they would both reduce the number of loading cycles experienced by reactor components, increasing their lifetime and improving the reactor duty cycle. The results of our negative triangularity scoping are summarized in Tbl.~\ref{tbl:negD} which lists the parameters of the discharge and Fig.~\ref{fig:negD} which shows the final POPCONs and GS equilibrium solution. Despite lower current, negative triangularity had approximately the same $q_{95}$ (reduction of the edge safety factor in negative triangularity has also observed in experiment \cite{Marinoni2021,Coda2021}), but the potential to greatly extend pulse time in negative triangularity merits further investigation of negative triangularity RPL-mode equilibria using gyrokinetic simulations.

\section{Conclusion and Future Work}

We described a novel operational scenario in high-field compact tokamak reactors, designated here as a radiative pulsed L-mode. In this operational scenario, an ARC-class tokamak is operated at both high current and high plasma density to overcome the low confinement present in L-mode to produce large amounts of fusion power. L-mode is maintained by the introduction of impurities into the plasma that radiate power before it may be convected through the scrape off layer. The RPL-mode allows us to avoid many of the operational concerns associated with H-mode, and perhaps most importantly, provides a core exhaust solution that allows tokamak power density to be maximized without the need for an advanced divertor. 

Using integrated modeling we have confirmed that a thermally stable, near-ignited, RPL-mode can be obtained. Our integrated modeling procedure is a major advance over the previous ARC studies, which used POPCONs with imposed profiles not self-consistent core transport simulations \cite{Sorbom2015,Kuang2018}. Additional analysis of RPL-modes using 'physics informed' POPCONs indicated the they will scale favorably to negative triangularity. Increased confinement from the reduction of turbulent transport lowers $I_p$, reducing $V_{loop}$ and extending the reactor's pulse time.  

Despite these promising initial results there are a number of methodological constraints present in our analysis here. Our model has considered only flattop operation, using a simplified magnetic field coil set, uniform impurity density profiles, and no sawtooth model (though we have accounted for the flattening of magnetic shear within the $q=1$ surface in our turbulent transport calculations). Once these constraints are addressed, optimization of plasma shaping and size with a focus on negative triangularity to reduce turbulent transport is a high-yield next step. This could be done using an optimization similar to that performed in ref. \cite{Highcock2018}. After transport optimization, startup modeling, field-coil-set optimization, and engineering design of an RPL-mode ARC reactor capable of accommodating the optimized plasma shape could be seriously attempted. Startup and current ramp modeling with more realistic coil-set and a sawtooth model will be required for accurate flux consumption estimation. It may be necessary to iterate between the plasma shaping optimization, startup, and coil-set modeling as coil positioning and currents as well as shaping can be closely linked to startup flux consumption through the external and plasma inductance. When accurate flux consumption calculations and a solenoid engineering design are obtained, one could estimate pulse time, determine component lifetimes under cyclic loads, and assess economic viability. Special care will also need to be taken during startup modeling to ensure that power exhaust can be satisfied throughout the current ramp while avoiding radiative collapse.

Finally, while RPL-mode ARC reactors eliminate many challenges present in AT reactor scenarios they have unique engineering challenges of their own. Disruptions are an especially serious concern in RPL-mode scenarios. The high plasma current densities and large radiated power fractions can trigger instabilities and radiative collapse. Both predictive solutions to disruptions, such as machine learning aided disruption avoidance \cite{Kates-Harbeck2019,Montes2019,Pau2019}, \textit{and} engineered solutions, such as advanced vacuum vessels \cite{Peterson2021}, passive coils \cite{Weisberg2021}, and innovative tearing mode suppression systems like the LHCD current condensation scheme investigated here, will be required in any future RPL-mode reactors. Furthermore, while peak power fluxes on plasma facing components will generally be lower in RPL-modes than other reactor scenarios there will be a substantial $\sim 1$ MW/m$^2$ radiative power flux to all plasma facing components which will require an innovative first wall design. If these challenges can be overcome, however, RPL-mode ARC reactors offer a path to some of the highest power-densities of any proposed fusion reactor.

\section*{Acknowledgements}
This work originated from the MIT Department of Nuclear Science and Engineering fusion reactor design course. The authors would like to thank all of the instructional staff from the course and give a special thanks to S.E. Ferry for her generous help editing this work. D.G. Whyte and J.P. Freidberg acknowledge the support of the NSE Department and the PSFC. S.J. Frank was supported by US Department of Energy grant: DE-FG02-91ER54109. C.J. Perks was supported by US Department of Energy grant: DE-SC0014264. A.O. Nelson was supported by US Department of Energy grants: DE-SC0022270, DE-SC0022272, DE-SC0015480, and DE-SC0015878. T. Qian was supported by NSF GRFP grant: DGE-2039656. S. Jin was supported by DOE/NNSA–Cornell subaward: 23105-G0001-10010405 and US DOE grants: DE-AC02-09CH11466 and DE-SC0016072. A. Reiman was supported by US Department of Energy grant: DE-AC02-09CH11466. P. Rodriguez-Fernandez was supported by US Department of Energy grant: DE-SC0014264. This research used resources of the National Energy Research Scientific Computing Center, a DOE Office of Science User Facility supported by the Office of Science of the U.S. Department of Energy under Contract No. DE-AC02-05CH11231 using NERSC award FES-ERCAP0020035. 

\section*{References}
\bibliography{docbib}
\end{document}